
%
%
\newbox\hdbox%
\newcount\hdrows%
\newcount\multispancount%
\newcount\ncase%
\newcount\ncols
\newcount\nrows%
\newcount\nspan%
\newcount\ntemp%
\newdimen\hdsize%
\newdimen\newhdsize%
\newdimen\parasize%
\newdimen\spreadwidth%
\newdimen\thicksize%
\newdimen\thinsize%
\newdimen\tablewidth%
\newif\ifcentertables%
\newif\ifendsize%
\newif\iffirstrow%
\newif\iftableinfo%
\newtoks\dbt%
\newtoks\hdtks%
\newtoks\savetks%
\newtoks\tableLETtokens%
\newtoks\tabletokens%
\newtoks\widthspec%
%
%
%
%
\tableinfotrue%
\catcode`\@=11
%
%
\def\tstrut{\vrule height3.1ex depth1.2ex width0pt}%
\def\and{\char`\&}
\def\tablerule{\noalign{\hrule height\thinsize depth0pt}}%
\thicksize=1.5pt
\thinsize=0.6pt
\def\thickrule{\noalign{\hrule height\thicksize depth0pt}}%
\def\ctr#1{\hfil\ #1\hfil}%
%
%
%
%
\tablewidth=-\maxdimen%
\spreadwidth=-\maxdimen%
\def\tabskipglue{0pt plus 1fil minus 1fil}%
%
%
\centertablestrue%
%
%
%
%
\parasize=4in%
\gdef\ARGS{########}
\gdef\headerARGS{####}
\def\@mpersand{&}
{\catcode`\|=13
\gdef\letbarzero{\let|0}
\gdef\letbartab{\def|{&&}}%
\gdef\letvbbar{\let\vb|}%
}
{\catcode`\&=4
\def\ampskip{&\omit\hfil&}
\catcode`\&=13
\let&0
\xdef\letampskip{\def&{\ampskip}}%
\gdef\letnovbamp{\let\novb&\let\tab&}
}
\def\begintable{
   \begingroup%
   \catcode`\|=13\letbartab\letvbbar%
   \catcode`\&=13\letampskip\letnovbamp%
   \def\multispan##1{
      \omit \mscount##1%
      \multiply\mscount\tw@\advance\mscount\m@ne%
      \loop\ifnum\mscount>\@ne \sp@n\repeat%
   }
   \def\|{%
      &\omit\widevline&%
   }%
   \ruledtable
}
\long\def\ruledtable#1\endtable{%
%
%
%
   \offinterlineskip
   \tabskip 0pt
   \def\widevline{\vrule width\thicksize}
   \def\endrow{\@mpersand\omit\hfil\crnorm\@mpersand}%
   \def\crthick{\@mpersand\crnorm\thickrule\@mpersand}%
   \def\crthickneg##1{\@mpersand\crnorm\thickrule
          \noalign{{\skip0=##1\vskip-\skip0}}\@mpersand}%
   \def\crnorule{\@mpersand\crnorm\@mpersand}%
   \def\crnoruleneg##1{\@mpersand\crnorm
          \noalign{{\skip0=##1\vskip-\skip0}}\@mpersand}%
   \let\nr=\crnorule
   \def\endtable{\@mpersand\crnorm\thickrule}%
   \let\crnorm=\cr
%
%
   \edef\cr{\@mpersand\crnorm\tablerule\@mpersand}%
   \def\crneg##1{\@mpersand\crnorm\tablerule
          \noalign{{\skip0=##1\vskip-\skip0}}\@mpersand}%
   \let\ctneg=\crthickneg
   \let\nrneg=\crnoruleneg
   \the\tableLETtokens
%
%
   \tabletokens={&#1}
%
%
   \countROWS\tabletokens\into\nrows%
   \countCOLS\tabletokens\into\ncols%
%
%
   \advance\ncols by -1%
   \divide\ncols by 2%
   \advance\nrows by 1%
%
%
   \iftableinfo %
      \immediate\write16{[Nrows=\the\nrows, Ncols=\the\ncols]}%
   \fi%
%
%
   \ifcentertables
      \ifhmode \par\fi
      \line{
      \hss
   \else %
      \hbox{%
   \fi
      \vbox{%
         \makePREAMBLE{\the\ncols}
         \edef\next{\preamble}
         \let\preamble=\next
         \makeTABLE{\preamble}{\tabletokens}
      }
      \ifcentertables \hss}\else }\fi
   \endgroup
   \tablewidth=-\maxdimen
   \spreadwidth=-\maxdimen
}
\def\makeTABLE#1#2{
   {
   \let\ifmath0
   \let\header0
   \let\multispan0
%
%
   \ncase=0%
   \ifdim\tablewidth>-\maxdimen \ncase=1\fi%
   \ifdim\spreadwidth>-\maxdimen \ncase=2\fi%
   \relax
%
   \ifcase\ncase %
      \widthspec={}%
   \or %
      \widthspec=\expandafter{\expandafter t\expandafter o%
                 \the\tablewidth}%
   \else %
      \widthspec=\expandafter{\expandafter s\expandafter p\expandafter r%
                 \expandafter e\expandafter a\expandafter d%
                 \the\spreadwidth}%
   \fi %
   \xdef\next{
      \halign\the\widthspec{%
      #1
      \noalign{\hrule height\thicksize depth0pt}
      \the#2\endtable
%
      }
   }
   }
   \next
}
\def\makePREAMBLE#1{
   \ncols=#1
   \begingroup
   \let\ARGS=0
   \edef\xtp{\widevline\ARGS\tabskip\tabskipglue%
   &\ctr{\ARGS}\tstrut}
   \advance\ncols by -1
   \loop
      \ifnum\ncols>0 %
      \advance\ncols by -1%
      \edef\xtp{\xtp&\vrule width\thinsize\ARGS&\ctr{\ARGS}}%
   \repeat
   \xdef\preamble{\xtp&\widevline\ARGS\tabskip0pt%
   \crnorm}
   \endgroup
}
\def\countROWS#1\into#2{
   \let\countREGISTER=#2%
   \countREGISTER=0%
   \expandafter\ROWcount\the#1\endcount%
}%
\def\ROWcount{%
   \afterassignment\subROWcount\let\next= %
}%
\def\subROWcount{%
   \ifx\next\endcount %
      \let\next=\relax%
   \else%
      \ncase=0%
      \ifx\next\cr %
         \global\advance\countREGISTER by 1%
         \ncase=0%
      \fi%
      \ifx\next\endrow %
         \global\advance\countREGISTER by 1%
         \ncase=0%
      \fi%
      \ifx\next\crthick %
         \global\advance\countREGISTER by 1%
         \ncase=0%
      \fi%
      \ifx\next\crnorule %
         \global\advance\countREGISTER by 1%
         \ncase=0%
      \fi%
      \ifx\next\crthickneg %
         \global\advance\countREGISTER by 1%
         \ncase=0%
      \fi%
      \ifx\next\crnoruleneg %
         \global\advance\countREGISTER by 1%
         \ncase=0%
      \fi%
      \ifx\next\crneg %
         \global\advance\countREGISTER by 1%
         \ncase=0%
      \fi%
      \ifx\next\header %
         \ncase=1%
      \fi%
      \relax%
      \ifcase\ncase %
         \let\next\ROWcount%
      \or %
         \let\next\argROWskip%
      \else %
      \fi%
   \fi%
   \next%
}
\def\counthdROWS#1\into#2{%
\dvr{10}%
   \let\countREGISTER=#2%
   \countREGISTER=0%
\dvr{11}%
\dvr{13}%
   \expandafter\hdROWcount\the#1\endcount%
\dvr{12}%
}%
\def\hdROWcount{%
   \afterassignment\subhdROWcount\let\next= %
}%
\def\subhdROWcount{%
   \ifx\next\endcount %
      \let\next=\relax%
   \else%
      \ncase=0%
      \ifx\next\cr %
         \global\advance\countREGISTER by 1%
         \ncase=0%
      \fi%
      \ifx\next\endrow %
         \global\advance\countREGISTER by 1%
         \ncase=0%
      \fi%
      \ifx\next\crthick %
         \global\advance\countREGISTER by 1%
         \ncase=0%
      \fi%
      \ifx\next\crnorule %
         \global\advance\countREGISTER by 1%
         \ncase=0%
      \fi%
      \ifx\next\header %
         \ncase=1%
      \fi%
\relax%
      \ifcase\ncase %
         \let\next\hdROWcount%
      \or%
         \let\next\arghdROWskip%
      \else %
      \fi%
   \fi%
   \next%
}%
{\catcode`\|=13\letbartab
\gdef\countCOLS#1\into#2{%
   \let\countREGISTER=#2%
   \global\countREGISTER=0%
   \global\multispancount=0%
   \global\firstrowtrue
   \expandafter\COLcount\the#1\endcount%
   \global\advance\countREGISTER by 3%
   \global\advance\countREGISTER by -\multispancount
}%
\gdef\COLcount{%
   \afterassignment\subCOLcount\let\next= %
}%
{\catcode`\&=13%
\gdef\subCOLcount{%
   \ifx\next\endcount %
      \let\next=\relax%
   \else%
      \ncase=0%
      \iffirstrow
         \ifx\next& %
            \global\advance\countREGISTER by 2%
            \ncase=0%
         \fi%
         \ifx\next\span %
            \global\advance\countREGISTER by 1%
            \ncase=0%
         \fi%
         \ifx\next| %
            \global\advance\countREGISTER by 2%
            \ncase=0%
         \fi
         \ifx\next\|
            \global\advance\countREGISTER by 2%
            \ncase=0%
         \fi
         \ifx\next\multispan
            \ncase=1%
            \global\advance\multispancount by 1%
         \fi
         \ifx\next\header
            \ncase=2%
         \fi
         \ifx\next\cr       \global\firstrowfalse \fi
         \ifx\next\endrow   \global\firstrowfalse \fi
         \ifx\next\crthick  \global\firstrowfalse \fi
         \ifx\next\crnorule \global\firstrowfalse \fi
         \ifx\next\crnoruleneg \global\firstrowfalse \fi
         \ifx\next\crthickneg  \global\firstrowfalse \fi
         \ifx\next\crneg       \global\firstrowfalse \fi
      \fi
\relax
      \ifcase\ncase %
         \let\next\COLcount%
      \or %
         \let\next\spancount%
      \or %
         \let\next\argCOLskip%
      \else %
      \fi %
   \fi%
   \next%
}%
\gdef\argROWskip#1{%
   \let\next\ROWcount \next%
}
\gdef\arghdROWskip#1{%
   \let\next\ROWcount \next%
}
\gdef\argCOLskip#1{%
   \let\next\COLcount \next%
}
}
}
\def\spancount#1{
   \nspan=#1\multiply\nspan by 2\advance\nspan by -1%
   \global\advance \countREGISTER by \nspan
   \let\next\COLcount \next}%
\def\dvr#1{\relax}%
\def\header#1{%
\dvr{1}{\let\cr=\@mpersand%
\hdtks={#1}%
\counthdROWS\hdtks\into\hdrows%
\advance\hdrows by 1%
\ifnum\hdrows=0 \hdrows=1 \fi%
\dvr{5}\makehdPREAMBLE{\the\hdrows}%
\dvr{6}\getHDdimen{#1}%
{\parindent=0pt\hsize=\hdsize{\let\ifmath0%
\xdef\next{\valign{\headerpreamble #1\crnorm}}}\dvr{7}\next\dvr{8}%
}%
}\dvr{2}}
\def\makehdPREAMBLE#1{
\dvr{3}%
\hdrows=#1
{
\let\headerARGS=0%
\let\cr=\crnorm%
\edef\xtp{\vfil\hfil\hbox{\headerARGS}\hfil\vfil}%
\advance\hdrows by -1
\loop
\ifnum\hdrows>0%
\advance\hdrows by -1%
\edef\xtp{\xtp&\vfil\hfil\hbox{\headerARGS}\hfil\vfil}%
\repeat%
\xdef\headerpreamble{\xtp\crcr}%
}
\dvr{4}}
\def\getHDdimen#1{%
\hdsize=0pt%
\getsize#1\cr\end\cr%
}
\def\getsize#1\cr{%
\endsizefalse\savetks={#1}%
\expandafter\lookend\the\savetks\cr%
\relax \ifendsize \let\next\relax \else%
\setbox\hdbox=\hbox{#1}\newhdsize=1.0\wd\hdbox%
\ifdim\newhdsize>\hdsize \hdsize=\newhdsize \fi%
\let\next\getsize \fi%
\next%
}%
\def\lookend{\afterassignment\sublookend\let\looknext= }%
\def\sublookend{\relax%
\ifx\looknext\cr %
\let\looknext\relax \else %
   \relax
   \ifx\looknext\end \global\endsizetrue \fi%
   \let\looknext=\lookend%
    \fi \looknext%
}%
%
%
\def\tablelet#1{%
   \tableLETtokens=\expandafter{\the\tableLETtokens #1}%
}%
\catcode`\@=12
%
\def\bs{\bigskip}

\def\cl{\centerline}

\def\i{\item}

\def\ni{\noindent}

\def\ub{\underbar}
\def\vs{\vskip}
\def\today{\ifcase\month\or
        January\or February\or March\or April\or May\or June\or
        July\or August\or September\or October\or November\or December\fi
        \space\number\day, \number\year}

\magnification=1200
\hsize=5.8truein
\vsize=8.5truein
\baselineskip=16pt
\parskip=0pt


\begingroup
\baselineskip=12pt

\cl{BILINEAR QUANTUM MONTE CARLO:}
\vs .1truein
\cl{EXPECTATIONS AND ENERGY DIFFERENCES}
\vs .5truein
\cl{Shiwei Zhang and M.H. Kalos}
\vs .2truein
\cl{Laboratory of Atomic and Solid State Physics}
\cl{and}
\cl{Center for Theory and Simulation in Science and Engineering}
\vs .1truein
\cl{Cornell University}
\cl{Ithaca, New York 14853, USA}
\vs .6truein
\cl{ABSTRACT}
\vs .2truein
\begingroup
\baselineskip=16pt
{\narrower\smallskip{
We propose a bilinear sampling algorithm in Green's function \break Monte
Carlo for expectation values of operators that do not commute with the
Hamiltonian
and for differences between
eigenvalues of different Hamiltonians.
The integral representations of the Schroedinger equations
are transformed into two equations whose solution has the form $\psi_a(x)
t(x,y) \psi_b(y)$, where $\psi_a$ and $\psi_b$ are the wavefunctions for
the two related systems and
$t(x,y)$ is a kernel chosen to couple $x$ and $y$. The
Monte Carlo process, with random walkers on the enlarged configuration
space $x \otimes y$, solves these equations by generating
densities whose asymptotic form is the above bilinear distribution.
With
such a distribution, exact Monte Carlo estimators can be obtained for
the expectation values of quantum operators and for energy differences.
We present results of these
methods applied to several test problems, including a model integral equation,
and the hydrogen atom.
\bs
{\ni {\bf Key words}: algorithm; bilinear sampling; energy difference;
Green's function Monte Carlo; quantum expectations; random walk.}

\smallskip}}

\vfill
\eject

\cl{I.\ \ INTRODUCTION}
\medskip
Quantum Monte Carlo (QMC) methods have found their way into increasingly
many applications in the study of many-body systems. Among them, the Green's
function Monte Carlo$^{[1],[2],[3],[4]}$ (GFMC) has proved a very powerful way
to solve the Schroedinger equation in many dimensions.
Based on an iterative stochastic process, asymptotically it yields a Monte
Carlo (MC) representation of the ground state wavefunction of the
system.
For quantum operators whose eigenfunctions are the wavefunctions,
such as the Hamiltonian itself, only one MC sample
is needed together with an analytical trial wavefunction to evaluate exactly
their expectation values. Therefore, the GFMC method is capable of calculating
ground state energies and related quantities in a relatively
straightforward fashion. Indeed numerous highly accurate and effective
calculations have been performed along these lines for many very
different systems.

Due to the random and discrete nature of MC, however,
difficulties arise in calculating
expectations which involve two wavefunctions. One obvious example
is the calculation of expectation values
of quantum operators that do not commute with the Hamiltonian.
One way of calculating expectations with respect to $\psi^2$ is to
generate enough independent samples from $\psi$ so that, with
substantial probability, two lie within range of a known Green's
function$^{[5]}$. For a many-body system, this requires two very
large samples.
Another example is the computation of small energy
differences between two systems,
described by similar Hamiltonians
whose ground state eigenvalues differ slightly. The statistical variance
associated with the mean in
energies computed independently can be comparable to the
energy difference, causing the signal in the calculation
of the difference to be lost in noise.

Attempts have been made
to overcome these difficulties,
leading to success in specific cases.
But the proposed approaches all present certain limitations,
making such computations
either inexact or much more complicated than the energy calculation.
Among these, the ``extrapolation''
method$^{[1],[6]}$ is the most straightforward and
has been widely applied to obtain
expectation values for quantum operators that do not commute
with the Hamiltonian. It gives a
biased estimate to the expectation value, and the bias is often
hard to assess.
A trial wavefunction is required which
accurately describes the desired property of the system. Thus for
systems that are not well understood, this method will always be uncertain.
Other approaches$^{[7],[8],[9],[10],[11]}$ for expectation values
keep track of decedents
of walkers or employ
``side walks''. These methods are in principle asymptotically exact
and they have been successfully applied to certain problems
to obtain very accurate results for expectation values.
But they are often technically quite
delicate and asymptotically unstable
in the sense that increasing the length of the side walks
so as to decrease the bias leads to a reduction of signal to
noise ratio. In the limit of infinite side walks, the ratio is zero.
Furthermore, their efficiency or even success largely depends also
on the quality of
the guiding wavefunction, i.e., {\it a priori} knowledge of
the true wavefunction.
For the problem of energy differences, it is sometimes possible to
carefully arrange to correlate the two
random walks such that the
errors in energies would largely cancel$^{[12]}$. But unfortunately
this is also limited in its applicability.

Underlying the difficulties of these calculations
is the fact that they require highly
correlated configurations in the MC process
representing two related functions.
We propose in this paper a bilinear sampling algorithm in the GFMC
framework that achieves such a correlation naturally within a single random
walk. We show that it is possible to apply this to calculate expectation
values of quantum operators that do not
commute with the Hamiltonian as well as to compute energy differences.
As a new and very different approach, it results from
the integral representation of the
Schroedinger equations used by standard GFMC calculations for energies.
Instead of sampling a function linear in the unknown wavefunctions,
we transform the Schroedinger equations into a pair of integral equations
whose solutions are bilinear in the wavefunctions.
The random walk based on the new pair of equations
{\it directly}
samples the ground state of such solutions, namely,
${\psi_a}_0(x) t(x,y) {\psi_b}_0(y)$, where $a$ and $b$ label the two related
systems and ${\psi_a}_0$ and ${\psi_b}_0$ are
correspondingly their ground state wavefunctions.
(The labels $a$ and $b$ can be the same, in which case
the wavefunctions describe the same system.) The function
$t(x,y)$ is chosen to couple appropriately the configurations $x$ and $y$.
The object of the random walk is an ensemble of pairs of configurations
$(x,y)$ rather than individual configurations $x$.

As we shall show in Section II, the bilinear sampling method
yields asymptotic distributions of configurations $(x,y)$ which
can provide exact MC estimators for
the function ${\psi_a}_0(x) {\psi_b}_0(x)$.
When $a=b$, this is simply the square of the ground state wavefunction
for a system, and we thus have an exact way to compute ground
state expectation values of quantum operators.
If $a$ and $b$ are different, it permits a direct calculation
of the energy difference as $\Delta E=\break
<{\psi_a}_0|\Delta H|{\psi_b}_0>
/<{\psi_a}_0|{\psi_b}_0>$, where $\Delta H = H_a - H_b$ is the difference
in the Hamiltonians for the two systems.

As tests,
we have applied this new algorithm to a model problem to compute
various moments
and also to the ground states of the hydrogen atom and related systems.
The former is based on an
integral equation composed entirely of Gaussians. Because of its
simplicity and flexibility, it provides a very
transparent picture of various aspects of the problem
and enables us to study the algorithm from many different angles.
In the latter, we calculate expectation values
of various operators for the ground state of H
and also energy differences between the ground states of H and similar
systems with different potentials. This provides tests of the necessary
ingredients for the application of the new algorithm to
larger systems of the same class.

In Section II, we outline
the formalism of the bilinear sampling method and its numerical
implementation. Then the application to two model systems
is described in Section III.
In Section IV, we give a discussion of
the method and its further possibilities. Finally in the Appendix
the sampling techniques involved in the hydrogen calculation
are developed. These are in fact general techniques necessary
for a class of such calculations for atoms and molecules.

\bigskip
\cl{II.\ \ FORMALISM AND THE RANDOM WALK}
\medskip

As we shall discuss in Section IV, the bilinear sampling approach is by
no means limited to a certain type of Green's function. But for
simplicity in presentation, we use the original form of GFMC$^{[2]}$.
In atomic units, the Schroedinger equation for a many-electron
system can be written as
$$[-{1\over 2} \nabla^2 + V(x)] \psi(x) = E \psi(x), \eqno(1)$$
where $x$ is a $3M$ dimensional vector denoting the coordinates
of all $M$ electrons in 3D space, and the energy is negative
for any bound state. We will use the Green's function $g(x,z)$
for the operator $(-{1\over 2}\nabla^2 - E)$ in $3M$ dimensions.
The energy $E$ is unknown, but it can be either scaled away for
Coulomb systems$^{[13]}$ or obtained iteratively, which usually
converges very fast (see Section III.2.2).
We can then transform the Schroedinger equation (1) into the following
integral equation:
$$\psi(x)=\lambda \int g(x,z) w(z) \psi(z) dz. \eqno(2)$$
The Green's function $g$ here has an analytically known form
and $w=-V$.
Eq (2) has a set of solutions $\psi$ with different eigenvalues
$\lambda$, the lowest of which is $1$.
With an arbitrary initial function having non-zero
overlap with the ground state wavefunction, this equation can be iterated
to yield asymptotically the solution corresponding to the
lowest $\lambda$, or the ground state $\psi_0$. This forms the basis
of the GFMC method. In practice, the process is carried out
with the wavefunction in each iteration represented by
a generation of individual configurations, or random walkers.
The walkers live in configuration
space and they move from one point $z$ in this space to
another $x$ according to the probability distribution
function $g(x,z)$. The function $w$ in Eq (2) is treated as a
multiplicative weight or as a source of branching of walkers.

We consider two related systems described by Eq (2):
$$\eqalignno{
&\psi_a(x)=\lambda_a \int g_a(x,u) w_a(u) \enskip \psi_a(u) du&(3a)\cr
&\psi_b(y)=\lambda_b \int g_b(y,v) w_b(v) \enskip \psi_b(v) dv,&(3b)\cr}$$
where again the subscripts $a$ and $b$ denote the two systems respectively.
In order to obtain an asymptotic solution bilinear (rather than linear)
in the wavefunctions, we use a coupling function $t(x,y)$ and
multiply each equation in Eqs (3) by $t(x,y)$ and the wavefunction
for the other
system to arrive at the following pair
of equations
$$\psi_a(x)t(x,y)\psi_b(y)=\lambda_a \int
{t(x,y)g_a(x,u)\over
t(u,y)}w_a(u) \enskip \psi_a(u)t(u,y)\psi_b(y)du
\eqno(4a)$$
and
$$\psi_a(x)t(x,y)\psi_b(y)=\lambda_b \int
{t(x,y)g_b(y,v)\over
t(x,v)}w_b(v) \enskip \psi_a(x)t(x,v)\psi_b(v)dv.
\eqno(4b)$$
These hold for any positive coupling function $t$ in principle
and their solution now has the form
$\psi_a(x)t(x,y)\psi_b(y)$. Eqs (4) are completely defined once the kernel
$t$ is chosen.
In this paper, we assume that $t$ is
symmetric in $x$ and $y$ solely for simplicity.
Eqs (4) can be rewritten so that they are more transparent for a random
walk interpretation:
$$\Phi(x,y)=\lambda_a \int
\Gamma_a(x,y|u,v) N_a(u,v)  \enskip \Phi(u,v) du dv
\eqno(5a)$$
and
$$\Phi(x,y)=\lambda_b \int
\Gamma_b(y,x|v,u) N_b(v,u)  \enskip \Phi(u,v) du dv,
\eqno(5b)$$
where $\Phi(x,y)=\psi_a(x)t(x,y)\psi_b(y)$ is the bilinear solution
we seek. The kernel $\Gamma_s$ is a normalized probability
distribution function of $x$ and $y$ conditional on $u$ and $v$ defined as
$$\Gamma_s(x,y|u,v)\propto [g_s(x,u) t(x,v)] \delta(y-v), \eqno(6)$$
where $s$ is either $a$ or $b$.
The multiplicative factor $N_s$ is
$$N_s(u,v)=w_s(u) {\int g_s(x,u) t(x,v) dx} /t(u,v). \eqno(7)$$
We note that neither $\Gamma_s$ nor $N_s$ is symmetric in its variables.

To solve Eqs (5) by Monte Carlo for the ground states in the bilinear form
$\Phi_0={\psi_a}_0(x) t(x,y) {\psi_b}_0(y)$,
we introduce random walks on the
enlarged configuration space $x\otimes y$. In other words, each walker
now consists of two configurations, namely $x$ and $y$, which sample a
joint distribution function
$\Phi(x,y)$. The iteration, or the random walk, is carried out
by moving walkers according to
either one of the two equations in each step. The initial distribution
of walkers can be arbitrary or generated from a Metroplis sampling of
$\Phi$ with the $\psi$'s replaced by trial wavefunctions for the ground
states.
The bilinear sampling technique enables us to treat explicitly functions
bilinear in the wavefunctions and maintain highly correlated configurations.
But since Eqs (5) are a direct transformation of Eqs (3),
the convergence property of the original linear integral equations
to the corresponding ground states simply carries through.
In the random walk process,
both equations need to be applied about equally often in order to
ensure convergence
and assure efficiency. In practice, this can be accomplished by using
the two equations alternately, or randomly with equal probability.
Once an equation has been selected in a step, the multiplicative factor
(either $N_a(u,v)$ or $N_b(v,u)$ as determined by the
equation) is constructed for each walker.
New walkers for the next generation are then produced by:
1) choosing a walker $(u,v)$ from the old population either
with probability proportional to their multiplicative factors
or by branching (depending on whether the size of the population
is fixed or not);
and 2) sampling a new walker from the parent walker
according to the kernel $\Gamma_s$ for that equation.
We note from the form of $\Gamma_s$ that
in fact in 2) only one new configuration $(x\ {\rm or}\ y$) is
selected and the complementary configuration in the old walker
is simply carried along.

Much freedom still remains about the choice of the coupling
function $t$.
If we set $t$ to a constant, bilinear sampling would
reduce to generating two independent sets
of configurations from two standard GFMC runs, from which the
overlap would be difficult to extract in high dimensional
systems.
The kernel $t$ must generate configurations that
are close together. The requirement of this is very clear in
the current problem. When $t$ is the same as one of the Green's functions,
say, $g_a$,
the bilinear sampling method yields a density
${\psi_a}_0(x) g_a(x,y) {\psi_b}_0(y)$. From the original integral equation,
Eq (3a), we have
$${\psi_a}_0(y){\psi_b}_0(y)={\lambda_a}_0 \int {\psi_a}_0(x) g_a(x,y)
{\psi_b}_0(y) w_a(x) dx. \eqno(8)$$
As mentioned in Section I, the product on the left-hand side of Eq (8)
is essential to the expressions for the expectation values we seek.
In the asymptotic regime of the MC process, the distribution of pairs of
configurations represents $\Phi_0(x,y)$, as a sum of delta
functions in $x$ and $y$:
$${\psi_a}_0(x) g_a(x,y) {\psi_b}_0(y)=\sum_k \delta(x-x_k) \delta(y-y_k),
\eqno(9)$$
where $\{x_k,y_k\}$ is the collection of these pairs (random walkers)
labeled by $k$.
This combined with Eq (8) gives,
$${\psi_a}_0(y){\psi_b}_0(y) \propto \sum_k w(x_k) \delta(y-y_k).
\eqno(10)$$
In other words, $w_a(x_k)$ is an MC
estimator for ${\psi_a}_0(y_k){\psi_b}_0(y_k)$.

With the direct sampling of the product of two wavefunctions
given by Eq (10), we can easily obtain the desired expectation
values of quantum operators exactly.
If $a=b$, the ground state expectation
value of any multiplicative quantum operator $O$ is
$$<O> \equiv {\int {\psi_a}_0(y) O(y) {\psi_a}_0(y) dy \over
\int {\psi_a}_0(y) {\psi_a}_0(y) dy}
={\sum_k O(y_k) w_a(x_k) \over \sum_k w_a(x_k)}. \eqno(11)$$
When $a \ne b$, we wish to calculate the ground state energy difference
between two systems $a$ and $b$ described by different
but related Hamiltonians $H_a$ and $H_b$. Their ground
state wavefunctions are given by ${\psi_a}_0$ and ${\psi_b}_0$, which
are assumed to be non-orthogonal to each other.
If $H_a - H_b =V_a - V_b$, since
$$\eqalignno{
\Delta E =& {\int {\psi_a}_0(y) H_a {\psi_b}_0(y) dy \over
\int {\psi_a}_0(y) {\psi_b}_0(y) dy}
-{\int {\psi_a}_0(y) H_b {\psi_b}_0(y) dy \over
\int {\psi_a}_0(y) {\psi_b}_0(y) dy} \cr
= & {\int {\psi_a}_0(y) [H_a - H_b] {\psi_b}_0(y)  \over
\int {\psi_a}_0(y) {\psi_b}_0(y) dy},  &(12)\cr}$$
we have
$$\Delta E =
{\sum_k w_a(x_k) [V_a(y_k)-V_b(y_k)] \over
\sum_k w_a(x_k)}. \eqno(13)$$
Of course, in both cases we can use Eq (3b) in Eq (8) to obtain
$w_b(y_k) {g_b(x_k,y_k) \over g_a(x_k,y_k)}$ as an estimator for
${\psi_a}_0(x_k){\psi_b}_0(x_k)$, which can provide expressions
similar to Eqs (11) and (13). They can be combined with the equations
shown here for better statistics.

Implicit above is the assumption that it is possible to
evaluate $N$ as well as to sample $\Gamma$. This is, of course,
not universally valid.
Fortunately, in the current GFMC approach, the Green's function
for $(-{1\over 2}\nabla^2 - E)$ for any system can be written
as a superposition of Gaussians and it is straightforward to
sample product of two such functions. The integrals in $N$ can
also be easily obtained.
The assumption is also true for other classes of Green's
functions for various systems of interest. For discrete systems
such as certain quantum spin systems,
it will be even less challenging in principle.

\bigskip
\cl{III.\ \ APPLICATIONS}
\medskip

\ni
{III.1. A Model Problem}
\medskip

In this part, we apply the bilinear sampling
method to a model problem$^{[14]}$ described
by an integral equation of exactly the same form as the
general many-dimensional equation given by Eq (2).
We shall test the capability of the algorithm in calculating
expectation values by evaluating various moments of the ``ground state''
distribution. The labels $a$ and $b$ (thus $s$) may be omitted
in this section.
Without altering the notation, we simply redefine as follows
$$g(x,z)={1 \over \sqrt{\alpha \pi}} {\rm exp}[- \alpha (x-z)^2]$$
and
$$w(z)={\sqrt{2\alpha \over 2\alpha-1}}
{\rm exp}(-{1 \over 4 \alpha-2} z^2),$$
where $\alpha$ can be any real number greater than $0.5$.
It can be easily verified that
under this new definition, Eq (2) has a set of solutions which are
product of Gaussians and polynomials. Among them the ground state,
the solution corresponding to the lowest eigenvalue $\lambda=1$,
is
$$\psi_0(x)={\rm exp}(-{1 \over 2} x^2).$$
We choose the coupling function $t$ to be also a Gaussian,
$$t(x,y)={\rm exp}[- \beta (x-y)^2],$$
and we shall solve Eqs (5) by MC
for the joint distribution $\Phi_0$ of the ground state.

With every term in Gaussian form, this problem is
easy to study analytically from every aspect.
The parameters $\alpha$ and $\beta$ are completely at our disposal
and they can be varied to provide insight into the behavior
of the algorithm under very different conditions of the Green's
function and coupling. Furthermore, if necessary, the
iteration can also be carried out directly without doing MC.
That is,
assuming a general solution ${\rm exp}(-a_n x^2 +b_n x y -c_n y^2)$
for the $n^{\rm th}$ iterate of Eqs (5), we can determine
$a_{n+1}$, $b_{n+1}$, and $c_{n+1}$ at every stage.
This can be used to generate numerical trajectories so as to observe
the convergence.
The product $g(x,u)t(x,v)$ in kernel\enskip $\Gamma(x,y|u,v)$
can be easily transformed into a single Gaussian in the unknown
$x$ by completing the square in the exponents. Not surprisingly, then,
sampling
$\Gamma$ amounts to sampling a Gaussian.

The MC process generates random walkers $\{ x_k,y_k \}$
representing
the distribution $\psi_0(x) t(x,y) \psi_0(y)$ for the ground state.
Since
$$\int \psi_0(x) t(x,y) \psi_0(y) {\rm exp}(-{1 \over 4 \beta-2} y^2) dy
\propto \psi_0^2(x)={\rm exp}(-x^2),$$
the MC result is adequate to determine completely the function
${\rm exp}(-x^2)$. For example, various moments
can be computed in a similar fashion as in Eq (11)
and compared with the exact results.
Also a histogram can be easily made for this function in one or two
dimensions.
The second moment $<x^2>$
is always computed in our tests. We have carried out calculations
in nine dimensions and the result is satisfactory. For
the present study, however, one dimension suffices in revealing
the characteristics of the algorithm.
A wide range of values have been used for both $\alpha$
and $\beta$. Table I presents our results for the second moment in one
dimension. We see that all the computed
answers agree well with the exact result, namely $0.5$.

One important issue is
how the algorithm performs when $\alpha$ and $\beta$ are
large, since that is the case when both kernels are very sharply
peaked. That the Green's function is sharp implies the step size
of the random walk is small in configuration space, which more
closely resembles the situation in higher dimensions resulting from actual
many-dimensional problems
and also the situation in which a QMC is generated by a diffusion process.
The coupling is consequently also
very sharp, as indicated in Section II. In fact in many cases such
as the hydrogen atom below, the coupling is simply the Green's function.
Under this circumstance, the probability becomes extremely small to have
two independent and random configurations appear close in configuration
space. Therefore the bilinear sampling method must effectively
couple two sets of configurations without distorting the
distribution. It is reasonable to expect less efficiency as
$\alpha$ and $\beta$ increase, but the sampled distribution must nevertheless
be correct. From Table I, we see this is indeed the case.

We keep a constant number of walkers in our simulations. This results
in a bias, since generations with high multiplicity contribute less
than they should and vice versa. Branching can be introduced instead
of strict population control to avoid this. It is also possible
to correct for such a bias by carrying weights$^{[15]}$.
The sum of multiplicative factors over all walkers in
a generation labeled by $n$, ${\cal N}_n$, indicates
the total number of walkers the next generation
should include. Thus with the number of walkers fixed at $L$ for every
generation, each generation can be assigned a weight formed
by a product of ${\cal N}/L$ from a certain number of
previous generations, i.e.,
${\cal W}_n=\prod^m_{l=1} {\cal N}_{n-l}/L.$
The number of previous generations to be included, $m$, can be
tuned so that it is large enough to remove
the bias and yet no excessive fluctuation is introduced.
For this model problem, with a fairly small
number of walkers (usually $1000$), it requires less than
ten generations to correct for the bias.
It is observed that, without the correction, the bias effect is often
quite significant. As the kernels become sharper, the bias
becomes more and more serious. Moreover, it does not seem to always
exhibit a clear $1/L$ behavior as in many GFMC calculations.
The population control bias
can be attributed to fluctuation of the multiplicative
function $N$. In bilinear sampling, due to the extra
function $t$ inserted to couple two points, it is not
implausible to have relative large variation in $N$. For instance, in
this model problem,
$N(u,v) \propto {\rm exp}[\beta^2 (u-v)^2/(\alpha+\beta)] w(u)$.
In these calculations caution
must be exercised to ensure that the result is unbiased, in other words,
robust against population size.

We also mention in passing that even though Eqs (4) and
therefore Eqs (5) are true for any non-zero $t$, they are not
necessarily always well-behaved in a MC calculation.
To illustrate this, recall the variance$^{[16]}$ of the total weights
for a generation in the random walk is given by
$\int N(u,v)^2 \Phi(u,v) dudv.$
But when the coupling function
$t$ is {\it much sharper} than the Green's function $g$,
this expression diverges, which implies that the MC
sampling would not actually converge to a definite answer.
For the model problem, it is possible to determine
the range of $\beta$ for each $\alpha$ where infinite
variance can be expected. We have verified that when the
parameters are given in this range, the MC answer can disagree
with the exact one significantly.
The above analysis has employed no knowledge of the specific
form of the kernels and thus is general to bilinear sampling.
For the purpose of computing the expectation values
of quantum operators, however, it is always possible to avoid
a kernel $t$ in that regime. So this should not pose any problem.
As an additional probe of the diverging weights, we can monitor
the fluctuation of population sizes in
a calculation.

\medskip
\ni
{III.2. The Hydrogen Atom}
\medskip

As another test case, we use the algorithm to study the ground
state of the hydrogen atom. In the first part, expectation
values of operators with respect to the ground state are
computed. In the second, we study similar systems with different
potentials and evaluate the energy difference between
the ground states of the new and original systems.
It is assumed the systems are non-relativistic with the nucleus fixed.
Except for technical details,
implementation of the new
algorithm to address these problems directly follows the formalism
developed in Section II. Based on Eq (8), the natural choice of $t$
is $t=g_a$, where $g_a$ is the Green's function for the hydrogen
atom.

In general, the Green's function $g$ for an $M$-electron system as
defined in Section II can be
written in an integral representation, which is a superposition
of Gaussians with different widths.
There also exists an analytical expression for $g$ in terms of
polynomials and modified Bessel functions of the second kind
so that it can be conveniently evaluated. Using these
expressions, the integral
in $N$ can be easily computed for any $M$ and there exists an efficient
way to sample $x$ from the product of functions in the
same class, $g_a(x,u) g_b(x,v)$ (cf. Appendix).

\medskip
\ni
{III.2.1 \it Ground State Expectation Values}

Similarly to III.1, we can sample the square of the ground state
wavefunction and compute {\it directly} within a single
run expectation values of
multiplicative operators. The sampling techniques are only a special
case of that discussed in the Appendix.
In Table II, we show computed expectation values of the potential
energy $V$, the radial distance $|x|$,
$x^2$, and the square of the third component of the electron
coordinate $x^2_3$, together with the exact answers.
The number of walkers is typically 3000 and the bias is not noticeable.
Because of the simplicity of the sampling process and the direct
sampling of the product of wavefunctions, the code is fast
and the algorithm quite efficient.
We see that the agreement between the bilinear sampling and the
exact results is again excellent.

\medskip
\ni
{III.2.2 \it Energy Difference Calculations}

In this part, we describe tests of the capability of the bilinear sampling
method to calculate energy differences by considering systems
similar to the H atom but with different potentials. We will employ
bilinear sampling to compute the energy differences between the ground
states of such systems and the hydrogen atom.
The Green's functions corresponding to the two Hamiltonians
belong to the same class discussed in the Appendix, only with a possible
difference between multiplicative factors in their arguments that
is a function of the energy.
The difference in the potential terms causes $w_a$ to differ
from $w_b$ in Eqs (4).

The simplest way to obtain such a system is to add a perturbation
term to the original hydrogen Hamiltonian, i.e.,
$H_b=H_a - \gamma H^\prime$, where $H^\prime$ is a
multiplicative operator and $\gamma$ is a small coefficient.  From
Eq (13), we can compute
$\Delta E/\gamma$ as a function of $\gamma$, where $\Delta E$ is
the energy difference
between the ground states of the two systems.
When $\gamma=0$,
this in fact is exactly the bilinear approach of calculating
the expectation value of $H^\prime$ in the ground state of $H_a$.
Therefore, in a sense, all results in III.2.1 can be viewed
as special cases of these calculations. As $\gamma$ is increased,
the result should deviate from the ground state expectation of
$H^\prime$ and should
always give the exact energy difference divided by $\gamma$. From the MC
point of view, it implies the ability to compute the exact energy difference
for all ranges of the parameter $\gamma$.
Moreover, the effect of the small perturbation is generated with small
fluctuation.
In fact, the statistical error may decrease with $\gamma$.

We consider a perturbation $H^\prime=1/|x|$. This is just the
original Coulomb potential and the energy difference is
trivially obtained analytically. In Fig 1,
we plot the computed $\Delta E/\gamma$
for some values of $\gamma$ from bilinear sampling and compare them
with the exact result. The ground state energy of H
is $-0.5$ in atomic units. We see that the agreement is excellent.
For instance, in the case of $\gamma=0.003$ the bilinear sampling method easily
yields an energy
difference of $0.0030036(12)$, which would be extremely challenging,
if possible at all,
for an approach by independent MC calculations.

We next study a system described by the Hamiltonian
$H_b$ for the so-called Hulth\'en potential
$$V_{\rm Hul}(|x|)
=-V_0 {{\rm exp}(-\rho |x|) \over 1-{\rm exp}(-\rho |x|)},$$
where $V_0$ and $\rho$ are parameters.
This system can be
solved exactly$^{[17]}$ and its ground state
energy is given by ${E_b}_0=-(2 V_0/\rho-\rho)^2/8$ \enskip
\enskip ($\rho^2<2 V_0$).
Let $V_0=\rho$.
Then this potential behaves like the Coulomb potential
at small values of $|x|$ and approaches zero exponentially at large
distances.
By varying the parameter $\rho$, we can control how similar this system
is to the hydrogen atom and the energy difference between their ground
states can be calculated from the bilinear sampling method and compared
with the exact results. In Table III, we show the computed and analytical
results for $\Delta E$ for some values of $\rho$. Again extremely
accurate values are easily obtained
with bilinear sampling for small as well as large energy differences.
(Each number
corresponds to roughly one hour on an IBM RS6000 workstation.)

As mentioned above, the argument of the Green's functions scales with
the ground
state energies and have the form $g(k_s|x-y|)$, where $g$ is the
Green's function
for the operator $(-\nabla^2+1)$ and $k_s=\sqrt {2|{E_s}_0|}$\quad ($s=a,b$).
This does not pose any difficulty because we can obtain
iteratively and very quickly the correct values
for the two energies.
In fact, it is observed that, quite generally, the final result of a standard
GFMC calculation in this approach is rather insensitive to the initial
input of the energy. We have tested the effect of iterations from a
very poor starting value of the energy ${E_b}_0$
in our calculations with the $1/|x|$ perturbation and indeed, the convergence
is very fast. For example, if we use
${E_a}_0$ as initial value for both Green's functions,
then within one run, we
can obtain the first order correction to ${E_b}_0$. Even for
large $\gamma$ the effect of the uncertainty of an initial energy guess
becomes unnoticeable in the final answer in one to a few more iterations.
In Fig 2, we illustrate this by plotting
$\Delta E/\gamma$ as a function of iterations for
two large values of the coefficient $\gamma$
($0.1$ and $0.5$).

The number of walkers in these calculations is typically $2000$
and we correct for the bias with several previous generations
as described in Section III.1. We alternate the two equations in Eqs (5).
In the bias correction, we calculate the averages
of ${\cal N}_l$ separately for even and odd $l$ and
normalize each accordingly such that these multiplicity factors
are kept around unity.
\bigskip
\cl{IV.\ \ DISCUSSION}
\medskip

It is interesting to note that the manipulation of the original
integral equation (3a) to arrive at Eq (4a) is very similar
to an importance sampling \break
transformation$^{[1],[4]}$. In fact, if the
coupling kernel $t(x,z)$ is the same as $g_a(x,z)$,
we can use as importance
function the correct form, namely, the unknown
wavefunction ${\psi_a}_0$ in its integral
representation as given by Eq (3a). Since the MC process
in effect does the integral in this importance function, we
only use the integrand and also drop the potential term $w_a$ and
Eq (4a) ensues. This also suggests that the algorithm should be
rather efficient and, explains to some degree why no trial wavefunction
is needed in the method.

As a straightforward method that depends little on {\it a priori}
knowledge, bilinear sampling should find itself useful to
different quantum problems as the GFMC approach is employed
more and more to understand various many-body systems.
For example, it seems possible to
study with this method ground state properties of certain
quantum spin systems$^{[11]}$. A generalized
version of the ideas developed here with quadrilinear sampling
provides the possibility of calculating transition moments
between two quantum states, as opposed to a method using
side walks$^{[10]}$.

By sampling the product of wavefunctions, the bilinear
approach is more promising than dealing with
two independent sets of configurations$^{[5],[10]}$ from two GFMC
calculations.
We need to study more the behavior of the method
as $g$ and $t$ become very sharp so as to give a general
prescription for avoiding the large biases or fluctuations we have
seen in that limit.
In many-electron systems, we inevitably will encounter the
``sign'' problem in quantum MC$^{[3],[13]}$. We have not yet formulated
a bilinear sampling algorithm  mechanism
that also addresses that problem in an exact way. But it is possible
to use bilinear sampling within the fixed-node$^{[3],[18]}$ approximation.
With this, the method is clearly generalizable to many-electron
systems in which the random walk is either
generated by a diffusion approximation or by domain Green's function
methods.  In either case, there are three possible choices for $t(x,y)$.
We can use $g(x,y)$ as in the present work, ignoring the fact that a
small number of estimates will be negative (because $w(x) = -V(x)$ is
negative.)  We can couple the two configurations $x$ and $y$ using the
kernel by which new points are generated (itself a Green's function
either in a short time approximation or over some finite domain.)
Finally, it is possible, in principle to couple the walkers by the full
Green's function (unknown in advance, but generated by walks that would
go from $x$ to $y$) and modifying the coupling recursively using local
Green's functions. Investigation of these alternatives will be the subject
of future research.
%

The bilinear sampling algorithm derives directly from the
integral representation of the many-body Schroedinger equation
of the system. It appears to be quite natural for the problems
involving functions quadratic in ground state wavefunctions,
and for the calculation of energy differences.
\bigskip
\cl{ACKNOWLEDGEMENT}
\medskip

We thank G. V. Chester, C.J. Umrigar, T. MacFarland, and S.A. Vitiello
for helpful discussions.
The Cornell Theory Center is funded by the U.S. National Science Foundation, by
New York State, by IBM, and by Cornell University.
\bs
\bs

\vfill
\eject

\cl{REFERENCES}
\medskip
\i{1.}D.M. Ceperley and M.H. Kalos, in {\it Monte Carlo Methods in
Statistical \break Physics}, ed. by K. Binder (Springer Verlag, 1979).
\i{2.}M.H. Kalos, Phys. Rev. \ub{128}, 1791, (1962).
\i{3.}K.E. Schmidt and M.H. Kalos, in {\it Applications of the Monte Carlo
Method in Statistical Physics}, ed. by K. Binder (Springer Verlag, 1984).
\i{4.}M.H. Kalos, D. Levesque, and L. Verlet, Phys. Rev. \ub{A9}, 2178 (1974).
\i{5.}M.H. Kalos, J. Comp. Phys. \ub{2}, 257, (1967).
\i{6.}P.A. Whitlock, D.M. Ceperley, G.V. Chester, and M.H. Kalos,
Phys. Rev. B \ub{19}, 5598 (1979).
\i{7.}M.H. Kalos, Phys. Rev. A \ub{2}, 250, (1970);
K.S. Liu, M.H. Kalos, and G.V. Chester, Phys. Rev. A \ub{10}, 303, (1974).
\i{8.}P.J. Reynolds, R.N. Barnett, B.L. Hammond, R.M. Grimes, and
W.A. Lester, Jr., Int. J. Quant. Chem.  \ub{29}, 589 (1986);
P.J. Reynolds, R.N. Barnett, B.L. Hammond, and W.A. Lester, Jr.,
J. Stat. Phys. \ub{43}, 1017 (1986).
\i{9.}R.N. Barnett, P.J. Reynolds, and W.A. Lester, Jr.,
J. Comput. Phys. \ub{96}, 258 (1991).
\i{10}R.N. Barnett, P.J. Reynolds, and W.A. Lester, Jr.,
J. Chem. Phys. \ub{96}, 2141 (1992);
R.N. Barnett, P.J. Reynolds, and W.A. Lester, Jr., unpublished.
\i{11.}K.J. Runge and R.J. Runge, in {\it Quantum Simulations of Condensed
Matter
Phenomena}, ed. by J.~D.~Doll and J.~E.~Gubernatis (World Scientific, 1990);
K.J. Runge, Phys. Rev. B (to be published).
\i{12.}B.H. Wells, Chem. Phys. Lett. \ub{115}, 89 (1985);
C.A. Traynor and J.B. Anderson, Chem. Phys. Lett. \ub{147}, 389(1988).
\i{13.}S. Zhang and M.H. Kalos, Phys. Rev. Lett. \ub{67}, 3074, (1991).
\i{14.}M.H. Kalos, in {\it Computational Atomic and Nuclear
Physics},
ed. by
C. Bottcher, M.R. Strayer and J.B. McGrory (World Scientific, 1989).
\i{15.}M.~P.~Nightingale, in {\it Finite Size Scaling and
Numerical Simulation of Statistical Systems}, ed. by V.~Privman
(World Scientific, 1990).
\i{16.}M.~H.~Kalos and P.~A.~Whitlock, {\it Monte Carlo Methods}, (Wiley,
New York, 1986).
\i{17.}S. Fl\"ugge, {\it Practical Quantum Mechanics I}, p175,
(Springer-Verlag, Berlin, 1971).
\i{18.}J.B. Anderson, J. Chem. Phys. \ub{63}, 1499 (1975); J. Chem. Phys.
\ub{65}, 4121 (1976); J. Chem. Phys. \ub{73}, 3897 (1980).
\i{19.}M.H. Kalos and S. Zhang, in {\it Recent Progress
in Many-body Theories}, vol.\enskip3, ed. by C.E. Campbell and E. Krotscheck
(Plenum, in press). The basic idea of
bilinear sampling is discussed briefly there.
\vfill
\eject
\cl{APPENDIX: SAMPLING THE PRODUCT OF TWO GREEN'S FUNCTIONS}
\medskip

In this appendix, we shall complete the technical part associated
with the sampling of kernel $\Gamma_s$ in Eq (6) and the evaluation
of the multiplicative factor $N$ in Eq (7). These are general to Green's
functions for the operator $(- {1 \over 2}\nabla^2+|E|)$ with any number of
particles. As discussed in Section III.2.2, we can assume the
energy is known.

Again let $M$ be the number of particles in the
system we treat. Then the Green's function$^{[2],[19]}$ as
defined above is given by
$$g(x,z)=({k^2 \over 4 \pi})^{3M/2} {\int_0^\infty t^{-3M/2} {\rm exp}
(-t-{k^2 |x-z|^2 \over 4t})dt}, \eqno({\rm A}1)$$
where $k=\sqrt {2 |E|}$.
As already mentioned, $g$ also has a form
which can be used to evaluate the Green's function:
$$g(x,z)=({k^2 \over 2 \pi})^{3M/2} K_{3M/2-1}(k|x-z|)/(k|x-z|)^{3M/2-1},
\eqno({\rm A}2)$$
where $K_m$ is the modified Bessel function of the second kind.
In the bilinear sampling process we need to evaluate $N_s$
and sample $\Gamma_s$.
With $t(x,y)=g_a(x,y)$, both $N_s$ and $\Gamma_s$ involve
the product of two Green's functions of the same class.
It is necessary to sample such products as well as evaluate integrals
based on them.

We consider the product of two Green's functions
with $k_a$ and $k_b$. Let $\xi=(k_b^2-k_a^2)/2$ \enskip($\ge 0$)\enskip and
$\sigma=(k_b^2 + k_a^2)/2$. From Eq (A1),
by completing squares and changing variables, it is
straightforward to
obtain the following expression:
$$\eqalignno{
&g_b(x,y) g_a(x,z) \cr
\propto &{\int_0^\infty {\int_{-p}^p
T(x|y,z,\tau_1,\tau_2) {\rm exp}(-\xi q)
\enskip p^{-3M/2} {\rm exp}(-\sigma p-{(y-z)^2 \over 4p})
\enskip dq dp}}.  &({\rm A}3)\cr}$$
The new pair of variables $p$ and $q$ are
$$\eqalignno{
&p=\tau_1+\tau_2 \cr
&q=\tau_1-\tau_2 \cr}$$
and $T(x|y,z,\tau_1,\tau_2)$ is a normalized probability distribution function
of $x$ conditional on $\tau_{1,2}$ (or $p$ and $q$), $y$, and $z$:
$$T(x|y,z,\tau_1,\tau_2)=({\tau_1+\tau_2 \over 4 \pi \tau_1 \tau_2})^{3M/2}
{\rm exp}[-{\tau_1+\tau_2 \over 4 \tau_1 \tau_2}
(x-{y \tau_2+z \tau_1 \over \tau_2+\tau_1})^2]. \eqno({\rm A}4)$$

Now we describe the actual sampling and integrating of this product of
Green's functions $g_b(x,y) g_a(x,z)$ given the positions of the
parent walkers $y$ and $z$.
Since integration over $x$ in Eq (A3) simply removes $T$, we have the
following
$${\int g_b(x,y) g_a(x,z) dx}
\propto {1 \over \xi} [k_a^{3M-2} g_a(y,z)-k_b^{3M-2} g_b(y,z)].
\eqno({\rm A}5)$$
As a special case, in the limit $k_a=k_b$, i.e., when the two Green's
functions correspond to
the same system, the above expression reduces to:
$${\int g_a(x,y) g_a(x,z) dx}
\propto {1 \over 2 \pi} k_a^{3M-4} g^{3M/2-1}_a(y,z),
\eqno({\rm A}6)$$
where the superscript of $g$ indicates the Green's function is for
$3M/2-1$ dimensions rather than $3M/2$ for the original
functions.
To sample an $x$ from a probability distribution function proportional
to $g_b(x,y) g_a(x,z)$, we note that for any known
pair of $y$ and $z$, Eq (A3) can be written as
$$g_b(x,y) g_a(x,z) \propto
{\int {\int T(x|p,q) Q(q|p) P(p) dp} dq}, \eqno({\rm A}7)$$
where $T$ is the same as in Eq (A4) with the variables
changed from $\tau_1$, $\tau_2$ to $p$ and $q$. $Q$ is a normalized
probability distribution function of $q$ conditional on $p$ as
follows
$$Q(q|p)=\cases{\xi {\rm exp}(-\xi q) /[
{\rm exp}(\xi p)-{\rm exp}(-\xi p)], &if $|q| \le p$\cr
              0,&otherwise.\cr} \eqno({\rm A}8)$$
$P$ is a positive function of $p$ on $(0,\infty)$ which therefore
can be viewed as a probability density function
$$P(p) \propto {1-{\rm exp}(-2 \xi p) \over 2 \xi p}
p^{-3M/2+1} {\rm exp}(-k_a^2 p-{(y-z)^2 \over 4p}). \eqno({\rm A}9)$$
Thus to obtain $x$ according to the probability density given by
Eq (A3) or Eq (A7), we need to sample a $p$ from Eq (A9), then
a $q$ on $[-p,p]$ according to (A8)
and finally sample the Gaussian in
Eq (A4). Sampling of (A9) is elementary$^{[16]}$ and
can be accomplished by,
e.g., sampling the exponential distribution in $p$ in the last term
and then doing rejections.
The case $k_a = k_b$ is once again straightforward, since $Q$ becomes uniform
on $[-p,p]$ and the first term in $P(p)$ is simply $1$.
We also mention that analogous (though possibly less elegant)
methods will apply for any coupling kernel that can be written in the
form
$$t(x,y)={\int_0^\infty h(t) {\rm exp}(-{(x-y)^2 \over 4t}) dt} $$
for $h(t) \ge 0$.
\vfill
\eject

Table I.

\bigskip
\begintable
$\alpha$|  \multispan{3}\tstrut\hfil $\beta$  :  result  \hfil\cr
$0.6$   |$0.6$  : $0.5005(5)$ & $1.0$  :  $0.5001(4)$ & $3.0$  :  $0.501(6)$\nr
$1.0$   |$0.6$  : $0.5001(2)$ & $1.0$  :  $0.4998(2)$ & $3.0$  :  $0.503(5)$\nr
$3.0$   | $1.0$ : $0.4999(7)$ & $3.0$  :  $0.501(3)$  & $5.0$  :  $0.501(2)$\nr
$5.5$   | $1.0$ : $0.5003(9)$ & $3.0$  :  $0.500(4)$  & $5.5$  :  $0.502(3)$\nr
$10.5$  |$0.6$  : $0.500(1)$  & $3.0$  :  $0.501(3)$  & $10.0$ :
$0.504(5)$\endtable

\bigskip

\centerline{Table Caption}

\smallskip
\noindent Table I. Results of the bilinear sampling method applied to the
1D model problem. Shown is the second moment $<x^2>$
from the sampled distribution for $\psi^2(x)$. The exact answer is
$0.5$. The parameters $\alpha$ and $\beta$ give the sharpness of
the Gaussian kernels $g$ (the Green's function) and $t$ (the coupling)
respectively. The statistical errors in the results are in the
last digits and are indicated in parentheses.

\bs
\bs
\bs
\vfill
\eject

Table II.

\bigskip

\begintable
item    &  $<V>$     & $<\sqrt {x^2}>$    &   $<x^2>$     &    $<x_3^2>$ \cr
bilinear& $-1.001(2)$&  $1.500(3)$&  $3.000(9)$   &   $1.002(3)$ \nr
exact   & $-1.0$     &  $1.5$     &  $3.0$       &   $1.0$     \endtable

\bigskip

\centerline{Table Caption}

\smallskip
\noindent Table II. Results of the new algorithm applied to the ground state
of the hydrogen atom together with the exact answers. The items
are the expectation values of the potential, the radial distance,
the second moment, and the
$z$-component of the second moment. All quantities are in atomic units.
The statistical errors in the MC results are in the
last digits and are indicated in parentheses.

\vfill
\eject

Table III.

\bs

\begintable
$\rho$   |  exact      &  bilinear$-$exact\cr
$0.001$  | $0.00049987500$ & $-.00000000004(8)$ \cr
$0.0125$ | $0.00623047$ & $.00000000(1)$\cr
$0.03333$| $0.01652778$    & $.00000007(6)$ \cr
$0.08333$| $0.0407986$  & $-.0000001(4)$\cr
$0.4$    | $0.180000$ &  $-.000004(4)$        \endtable

\bs

\centerline{Table Caption}

\smallskip

\noindent Table III. Energy differences $\Delta E={E_b}_0 - {E_a}_0$
from bilinear
sampling between the ground states of the hydrogen atom
and a similar system described by the Hulth\'en potential, compared with
exact results. The Hulth\'en potential is given by
$V_{\rm Hul}(|x|)=-\rho {\rm exp}(-\rho |x|)/[1-{\rm exp}(-\rho |x|)],$
whereas the Coulomb potential in H is $V(|x|)=-1/|x|$.
Atomic units are used. The ground state
energy of H is ${E_a}_0=-0.5$ and ${E_b}_0$ is higher.
The first column is the exact result for $\Delta E$, while the second column
gives the error (bilinear$-$exact) in the bilinear result.
The statistical errors from MC are again in the last digits and are shown
in parentheses in the second column.

\bs
\bs
\bs
\bs
\bs
\bs
\bs

\vfill
\eject

\centerline{Figure Captions}
Fig. 1. The ``effective'' energy difference $\Delta E/\gamma$ between
the original and perturbed systems for a
perturbation $\gamma V(x)$ to the ground state of the hydrogen
atom. Results of the bilinear sampling method are shown with
statistical fluctuations and are compared with the exact result.
$V(x)$ is the potential energy operator.
The bilinear calculations were done with the arbitrarily chosen
values $\gamma=0.003,
0.005, 0.01, 0.05, 0.1$.
The first order
perturbation result is $1$. Atomic units are used.

\bs
Fig. 2. Convergence of iterations with the input of the energy ${E_b}_0$
of the perturbed system in the energy difference
calculations for H. The perturbation is again
$-\gamma/|x|$ and large perturbations
($\gamma=0.1$ and $0.5$) are chosen in order to show a visible
convergence process.
Values of $\Delta E/\gamma$ are plotted together with statistical errors
as a function of the number of iterations. The exact results are
given as straight lines.
In both calculations an initial value $0$ is assumed for the energy
difference. In one iteration,
they both give $\Delta E/\gamma=0.9995(9)$, equivalent to the first
order perturbation result.
The calculation with $\gamma=0.1$ (lower curve) requires only one more
iteration to converge to the exact value $1.05$ while
that with $\gamma=0.5$ converges in four more iterations
to the correct answer $\Delta E/\gamma=1.25$.

\bye